\documentclass[letterpaper,aps,twocolumn,superscriptaddress,showkeys]{revtex4}
\usepackage[utf8]{inputenc}
\usepackage{amsmath, graphicx, hyperref}
\hypersetup{colorlinks=true, linkcolor=blue, citecolor=blue, urlcolor=blue}

\begin{document}

\title{A simple model to study phylogeographies and speciation patterns in space}

\author{Philippe Desjardins-Proulx}
\email[E-mail: ]{philippe.d.proulx@gmail.com}
\affiliation{Universit\'e du Qu\'ebec \`a Rimouski, Canada.}
\affiliation{Universit\'e du Qu\'ebec \`a Montr\'eal, Canada.}
\affiliation{Quebec Center for Biodiversity Science, McGill University, Canada.}
\affiliation{College of Engineering, University of Illinois at Chicago, USA.}

\author{James L. Rosindell}
\affiliation{Imperial College London, United-Kingdom.}

\author{Timoth\'ee Poisot}
\affiliation{Universit\'e du Qu\'ebec \`a Rimouski, Canada.}
\affiliation{Quebec Center for Biodiversity Science, McGill University, Canada.}

\author{Dominique Gravel}
\affiliation{Universit\'e du Qu\'ebec \`a Rimouski, Canada.}
\affiliation{Quebec Center for Biodiversity Science, McGill University, Canada.}


\begin{abstract} In this working paper, we present a simple theoretical
framework based on network theory to study how speciation, the process by which
new species appear, shapes spatial patterns of diversity. We show that this
framework can be expanded to account for different types of networks and
interactions, and incorporates different modes of speciation.  \end{abstract}

\maketitle

\section{Motivation}

The peculiar spatial relationship between closely related species was among the
first patterns of diversity used to infer evolution. As early as the 1850s,
Alfred Wallace noted that the closest relatives were often observed in adjacent
yet non-overlapping regions \cite{wal52,wal55}. Wagner and Jordan later relied
on a similar observation to argue for the importance of geography and isolation
in the formation of new species \cite{coy04}. And finally, Mayr developed a
theory of allopatric speciation, a cornerstone of the modern synthesis, again
using similar observations \cite{may42,may63}. The relationship between
phylogeny and geography has shaped our understanding of the origin of species
\cite{coy04,los09b}. It is also crucial to the development of a unified theory
of community assembly \cite{ric08,mou12}. Yet, theory remains mostly silent
about the subject. Few models can generate phylogeographies, and none can be
used to study the effect of complex spatial structures \cite{bar00}. This is
surprising, not only because of the theoretical importance of phylogeography,
but also because several phylogenetic methods use geography to infer patterns of
speciation \cite{lyn89,bar00,los03}.

Part of the problem lies in the limitations of traditional mathematical methods:
analytical solutions to spatially-explicit models are often only available for
the most trivial cases \cite{epp10}. Thus, we are left with no theoretical
framework to study the patterns noted by Wallace, Wagner, and Mayr. In this
document, we describe a very simple algorithm to generate phylogeographies in
spatial networks. Our approach is inspired by metapopulation theory
\cite{lev69,han99,han09} although the spatio-temporal scale is different: we're
interested in the dynamics of populations at the regional scale during long
periods. The model will be used to study phylogeographies in various spatial
contexts and to develop better tools to understand the relationship between
phylogeny and geography.

We use the term ``phylogeography'' in the general sense: it is the union of
phylogenetics with geography. Our approach emphasizes how spatial patterns of
speciation shape biodiversity. It cannot be used to study within-species
variations, a major focus of phylogeography \cite{fre11b}. This is more
consistent with the field known as comparative phylogeography.

\section{Modeling the landscape}

We model the landscape as a spatial network of communities. A network is a
flexible mathematical object defined as a set of vertices $V$ and a set of edges
$E$, which are used to connect the vertices \cite{new10}. Here, the vertices
represent communities and the edges denote migration
\cite{eco08,eco10,phdp-likely,phdp-complex}.  Spatial networks are simply
networks in which vertices are embedded in a known topological space
\cite{kob94}, in our case a two-dimensional map. Thus, each community is
represented by a vertex in the network and to a position on a map.  Networks are
increasingly common in ecology as they can be used to model complex structures
and quantify the effect of clustering, connectivity, and isolation
\cite{min07,min08,urb09,dal10}. In particular, isolation is the most important
factor in many speciation events \cite{coy04}, making networks well-suited to
study patterns of speciation in different contexts
\cite{phdp-likely,phdp-complex}. The spatial network can be built in two ways.
First, random geometric networks can be generated by randomly placing the
vertices on a surface, normally the unit square, and linking all communities
within some threshold distance \cite{pen03}.  This technique is used to test
network algorithms applied to maps \cite{sed01}.  Second, a real map can be used
as a template for a spatial network \cite{min07,dal10}. This method offers the
opportunity to generate predictions specific to a given spatial structure, and
test the predictions of our algorithm against empirical data.

\begin{figure} \centering\includegraphics[width=0.45\textwidth]{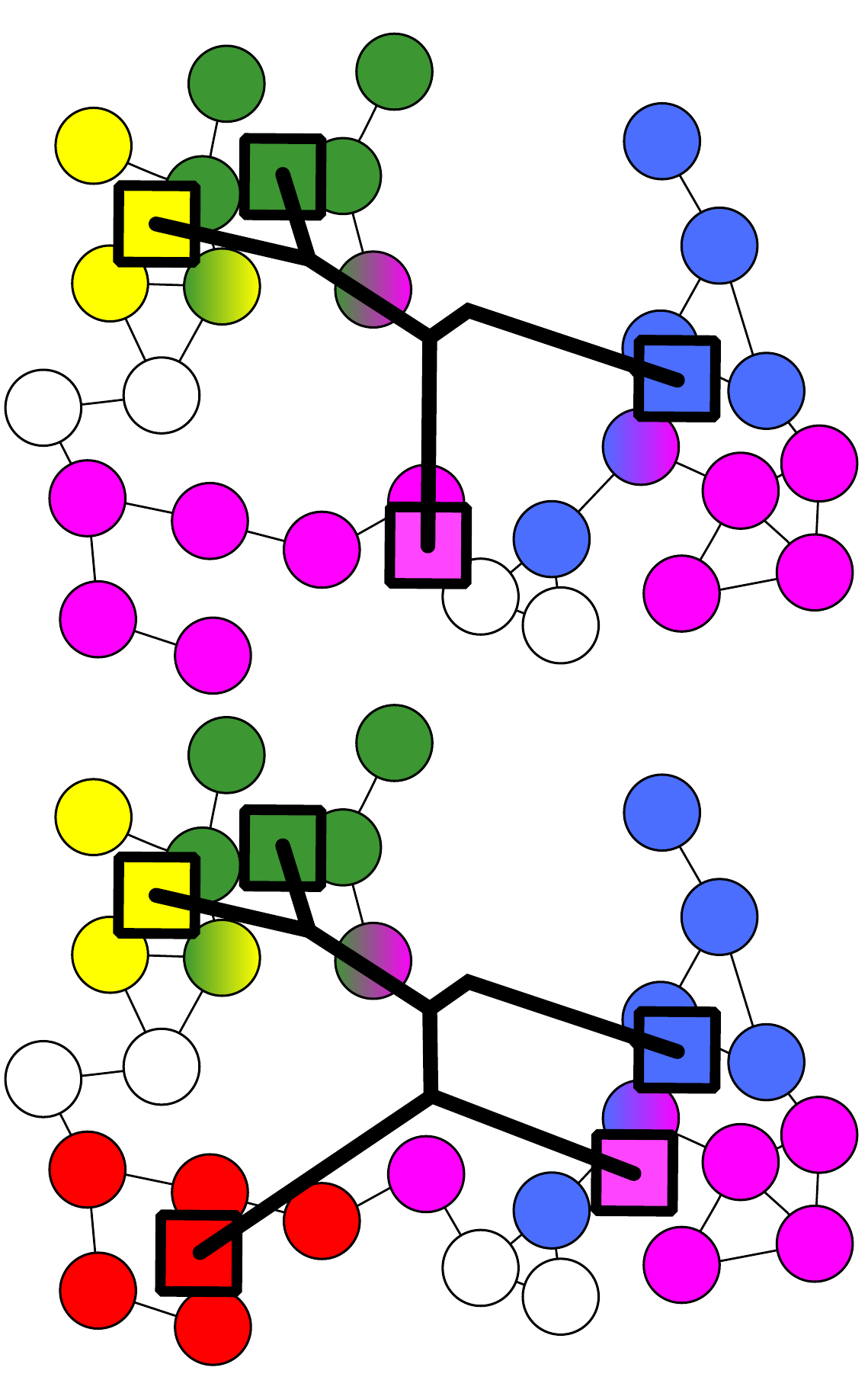}
\caption { Top: a phylogeography with four species (yellow, blue, green, pink).
The populations are distributed in a spatial network, with each community
(circles) hosting populations from 0 or more species. Empty communities are
white and a gradient is used for communities with more than one species. The
communities are connected by migration (thin black lines). Bottom: a speciation
event. The pink species is divided in three groups of populations. Its leftmost
group undergoes speciation and a connected subgroup now belongs to a new species
(in red).  } \label{fig:geophylo} \end{figure}

\section{The model}

A species is divided in populations which are distributed in a network of
communities. A species is either present or absent in a community, we do not
keep track of the number of individuals. Occupancy thus follows the standard
colonization/extinction dynamics of metapopulation theory \cite{han99}. For each
time step, all populations have the opportunity to colonize adjacent communities
(the vertices connected by an edge in the network). The probability of a
successful colonization of community $x$ by species $i$ is

\begin{equation}
   c(i,x) = c_{max}\exp\left(-\aleph\sum_{j \in \{S_x\backslash i\}} \delta_{ij}^{-1}\right),
\end{equation}

with $\{S_x \backslash\ i\}$ being the set of populations present in community
$x$ minus $i$, $\delta_{ij}$ is the time since species $i$ and $j$'s most recent
common ancestor, $c_{max}$ is the highest possible colonization rate and
$\aleph$ a positive constant (with $\aleph \geq 0$).  $\aleph$ describes the
decline of the intensity of interactions with phylogenetic divergence. In short,
a higher $\aleph$ makes it difficult for closely related species to coexist.
$c_{ix}$ is a very simple function derived from exponential decay. It is based
on an old hypothesis by Darwin: closely related species are more likely to
compete. It has recently received experimental support \cite{jia10,vio11}. A
strong assumption of trait conservatism underlies the model \cite{los08a}. At
each time step, all populations have the same probability $e$ of extinction.
Speciation occurs in groups of populations. We define a group as a set of
connected populations from the same species (Fig. \ref{fig:geophylo}). Each
group has a probability $v$ of undergoing speciation. When speciation occurs in
a group, a random subset of $[1, n]$ connected populations will speciate, with
$n$ being the number of populations in the original group (Fig.
\ref{fig:geophylo}).

\section{Variations}

The basic model can easily be extended to account for various types of
interactions. In this section we briefly discuss a few extensions.

\subsection{Allopatric speciation}

Our model is mostly parapatric, with strictly allopatric speciation occurring
only with probability $1/n$, with $n$ being the size of the group to speciate.
An alternative is to always force allopatric speciation by making the entire
group speciate.

\subsection{Sympatric speciation?}

With few solid cases of sympatric speciation, and many of them involving
important allopatric/parapatric phases \cite{coy04,bol07,fit08}, it is hard to
decide how to do a phenomenological sympatric speciation model. Furthermore, the
assumption of strong niche conservatism would be hard to maintain, as niche
overlap between diverging populations is one of the hardest challenge for
sympatric speciation. Nevertheless, if enough sympatric speciation events can be
analyzed, our model could be modified to allow sympatric, parapatric, and
allopatric speciation.

\subsection{Variable $\alpha$}

$\aleph$ is fixed in the original model, but it could vary in time and space.
For example, smaller regions could have higher $\aleph$ to account for a lower
carrying capacity. 

\subsection{Variable extinction rates}

The extinction rate could have the same form as the colonization rate and be
affected by closely related species.

\subsection{Variables $v$}

The speciation rate could decrease with higher diversity (more niches are
filled) or increase (``diversity begets diversity'') \cite{erw05}.

\subsection{Growing food webs}

The basic idea of using spatial networks and groups of connected populations for
speciation could be used to model how complex food webs grow with speciation
events. This integration would, however, require many new assumptions and a more
sophisticated model for the colonization and extinction rates.

Integrating food web dynamics lead to some difficulties. For example, a trophic
model would involve very different species with potentially different rates of
dispersal. The threshold value $r$ used to determine the realized links in the
spatial network would have to be different for each group of species. For example,
group-specific threshold values could be linked to the niche value (i.e.:
smaller species have lower dispersal ranges). A connected random geometric
networks could then be generated with the lowest threshold value, ensuring that
all networks are fully connected.

\subsection{Positive interactions}

Positive interactions between closely related species are also possible, for
example in plants. This variation can be achieved by making $c_{ix}$ increase
when related species are present.

\section{Implementation}

An open-source implementation is available on github:
\href{https://github.com/PhDP/wagner}{https://github.com/PhDP/wagner}.

\newpage
\bibliographystyle{plain}
\bibliography{../phdpdl}

\end{document}